\input jnl.tex
\singlespace
\preprintno{UF-IFT-HEP-94-16}
\preprintno{November 1994}
\vskip 2cm
\centerline{\bf ELECTROWEAK FLAVOR-CONSERVING}
\centerline{\bf GAUGE PROCESSES: VIRTUAL EFFECTS$^{*}$}
\vskip 2cm
\centerline{\bf Dallas C. Kennedy$^{\dagger}$}
\vskip .5cm
\centerline{\it Department of Physics, University of Florida}
\centerline{\it Gainesville, FL 32611 USA}
\vskip 2cm
\centerline{ABSTRACT}
\itemitem{}{\tenrm The electroweak Standard Model is summarized at the classical
and quantum levels, including its gauge symmetry and symmetry-breaking aspects.
The full implications of precise measurements of electroweak gauge forces are
presented in terms of electroweak parameters and quantum corrections.  The
minimal Standard Model (SM) (including the top quark) satisfies the data well, 
up to one-loop accuracy.  Possible non-Standard states subject to 
electroweak forces in quantum corrections is highly restricted by the present
data.  The status of exact and approximate symmetries of the electroweak
Standard Model is summarized.}
\footnote{}{$^*$~~Contribution to the American Physical Society/Division of
Particles and Fields Drell Panel Study of American High Energy Physics, Working
Subgroup 5.9: Electroweak Symmetry Breaking and Beyond the Standard Model:
Virtual Effects.}
\footnote{}{$^\dagger$~~e-mail: kennedy@phys.ufl.edu.}

\vfill\eject

\def\vev{\langle\phi\rangle}

{\twelvebf Experiments}
\bigskip
{\twelveit 1. Electromagnetic Fine Structure Constant $\alpha_{\rm em}$}
\bigskip
This parameter is the basic input to both quantum electrodynamics and the
electroweak Standard Model and
is best defined at energies well below electron-positron production
threshold, so that relativistic field theoretical effects are unimportant~[1].    
It can be measured in a variety of ways, including methods of
atomic, condensed matter, and low-energy scattering experiments.  The best
determinations of $\alpha_{\rm em}$ are made in macroscopic quantum Hall
effect or AC Josephson junction experiments, yielding~[2]:
$$\eqalign{
\alpha^{-1}_{\rm em} = 137.0359895\pm 0.0000061\quad ,}
$$\noindent
accurate enough for the purposes of high-energy physics measurements.
\bigskip
{\twelveit 2. Muon Decay and the Fermi Constant}
\bigskip
The fundamental constant characterizing the strength of the weak interactions
is the {\twelveit Fermi constant,} $G_\mu ,$ derived from the measured decay
lifetime of the muon in the beta process $\mu^-\rightarrow e^- \nu_\mu
{\bar\nu_e}~[3,4]:$
$$\eqalign{
\tau^{-1}_\mu = {G^2_\mu m^5_\mu\over{192\pi^3}}F(m^2_e/m^2_\mu )\big(1 + 
3m^2_\mu/5M^2_W\big)\big[
1 + {\alpha (m_\mu )\over{2\pi}}\big({25\over 4} -\pi^2\big)\big]\quad ,}
$$\noindent
where:
$$\eqalign{
F(x) = 1 - 8x + 8x^3 - x^4 - 12x^2\ln x\quad ,\cr
\alpha (m_\mu )^{-1} = \alpha^{-1} - {2\over 3\pi}\ln (m_\mu /m_e) +
{1\over 6\pi}\quad ,}
$$\noindent
with $\tau_\mu$ = (2.19703$\pm$0.00004)$\times$10$^{-6}$ sec, $G_\mu$ =
(1.16639$\pm$0.00002)$\times$10$^{-5}$ GeV$^{-2},$ and $m_\mu /m_e$ = 
206.768~[4].  This relation involves
only electromagnetic radiative corrections~[5].  From $G_\mu$ can be extracted
a universal weak decay constant $G_F,$ after the application of the remaining,
purely weak radiative corrections.  This constant $G_F$ is related to the 
Higgs vacuum expectation value (VEV):
$$\eqalign{
\langle\phi\rangle^2 = 1/\sqrt{2}G_F\quad ,}
$$\noindent
so that $\langle\phi\rangle\simeq$ 246 GeV sets the energy scale for the weak
interactions and the electroweak symmetry breaking.
\bigskip
{\twelveit 3. Deep Inelastic Neutrino Scattering}
\bigskip
The 1980s saw a series of $\nu N$ (neutrino-nucleon) deep inelastic scattering
(DIS) scattering experiments, carried out by the CDHS~[6] 
and CHARM~[7] collaborations
(CERN) and the CCFR collaboration (FNAL)~[8],
yielding measurements of $\sin^2\theta_W$ with accuracies to a few percent.
These measurements are conventionally quoted, for theoretical convenience,
in terms of the on-shell definition and are especially sensitive to the
top quark mass $m_t.$  The experiments are based on the scattering of
high-energy neutrino beams (100-200 GeV) off of fixed nuclear targets,
in the {\twelveit deep inelastic} regime; that is, the momentum transfer
is large (compared to the nucleon mass) and spacelike, and the nucleon is
transformed into a shower of hadrons.  In this regime, the neutrinos couple
directly to the underlying quarks (partons) of the nucleon, rather than to
the nucleon as a whole.  The scattering events are of two types, neutral
(NC) and charged current (CC).  The first is mediated by the $Z$ and leaves
the neutrino unchanged: the second is mediated by the $W$ and transforms the
neutrino into its charged lepton partner.  Since the neutrino beams are
obtained from muon decays, the (anti)neutrinos used are of the muon type,
$\nu_\mu$ or $\bar\nu_\mu$~[9,10].

The basic quantity of interest is the ratio of NC to CC $\nu_\mu$ events,
$R_\nu$ = $\sigma_{NC}(\nu )/\sigma_{CC}(\nu ).$  At the classical level,
this ratio, in terms of $\sin^2\theta_W,$ is:
$$\eqalign{
R_\nu = {1\over 2} - \sin^2\theta_W + {5\over 9}\sin^4\theta_W(1+r)\quad ,}
$$\noindent
where $r$ = $\sigma_{CC}(\bar\nu )/\sigma_{CC}(\nu )$ requires the use of
an antineutrino beam.  Although a nucleon is, at the level of valence quarks,
made up of only up $(u)$ and down $(d)$ quarks, it also contains a sea of
virtual quark-antiquark pairs of all types, suppressed by their masses.
Because of the presence of strange $(s)$ quarks and Cabibbo mixing between
the first and second families of quarks, the CC mode of $\nu N$ scattering
involves the production of heavy charm $(c)$ quarks; e.g.,
$$\eqalign{
\nu_\mu + d\rightarrow \mu^- + c\quad ,\quad
\nu_\mu + s\rightarrow \mu^- + c\quad ,}
$$\noindent
where the first is Cabibbo-suppressed and the second is the sea contribution.
The cross section
for this process involves the $c$ mass $m_c,$ thus introducing an uncertainty.
The charm-mass threshold effect is usually modelled using the {\twelveit slow
rescaling} method that introduces an unknown, effective $m_c.$ 

The CHARM collaboration of CERN measured $\nu N$ scattering, yielding a
result for $\sin^2\theta_W$ = 0.236+(0.012)($m_c -$ 1.5)$\pm$0.005(exp)
$\pm$0.003(th), with the effective $m_c$ = 1.5$\pm$0.3 in GeV~[7].  The CDHS
collaboration of CERN measured $\sin^2\theta_W$ = 0.225+(0.013)($m_c -$ 1.5)
$\pm$0.005(exp)$\pm$0.003(th), with the same $m_c$ derived from an earlier
CDHS experiment~[7].\footnote*{The values of $\sin^2\theta_W$ here are adjusted
for $m_t\simeq$ 180 GeV.}

The most recent reported result for neutrino-nucleon DIS is that of the
CCFR collaboration at Fermilab, with $\sin^2\theta_W$
= 0.2218$\pm$0.0025(stat)$\pm$0.0036(syst)$\pm$0.0040(th).  
The value of $m_c$ was measured, rather than assumed, from the observation
of dimuon events: $\nu_\mu + s\rightarrow\mu^- + c,$ then $c\rightarrow
s + \mu^+ + \nu_\mu .$  The effective $m_c$ = 1.31$\pm$0.24 GeV~[8].

The combined weighted average of all DIS experiments yields $\sin^2\theta_W$
= 0.2260$\pm$0.0048~[4].

The next neutrino-nucleon DIS experiment, probably the last significant
improvement in such measurements, will be the Fermilab E815 experiment
by the NuTeV collaboration, a descendant of CCFR, scheduled to begin in
late 1995, with preliminary results by early 1997.  Its goal is
the measurement of $\sin^2\theta_W$ to $\pm$0.0025, with the unique feature
of eliminating the charm threshold uncertainty.  The measured quantities
will be $R_{\pm}:$
$$\eqalign{
R_\pm &= {{\sigma_{NC}(\nu )\pm\sigma_{NC}(\bar\nu )}\over{\sigma_{CC}(\nu )\pm
\sigma_{CC}(\bar\nu )}}\cr
     &= {{R_\nu\pm rR_{\bar\nu}}\over{1\pm r}}\quad ,}
$$\noindent
using both neutrino and anti-neutrino beams.  The $m_c$ dependence of $r$
has been measured by the Fermilab experiments E774 and E770.  $R_-$ is
independent of charm production, depending only on valence quarks; with a
fixed $\rho ,$ $\sin^2\theta_W$ can be extracted from $R_-.$  (There is
a small charm production in $R_-$ from $d$ quarks, with Cabibbo suppression
$\sin^2\theta_C.$)  With $r$ known, $R_\nu$ and $R_{\bar\nu}$ can be 
inferred from $R_+,$ whence, combined with $R_-,$ independent values of
$\rho$ and $\sin^2\theta_W$ can be derived~[11].
\bigskip
{\twelveit 4. Neutrino-Electron Scattering}
\bigskip
The leptonic analogue to DIS is (anti)neutrino-electron scattering.  Two
older experiments of this type are those of the CHARM I (CERN)~[12] and 
E734 (BNL)~[13] collaborations.

The best and most recent of these measurements was that performed by the 
CHARM II collaboration (CERN)~[14].  The relevant quantities are the neutrino 
and antineutrino neutral current scattering cross sections and their ratio,
$R = \sigma (\nu_\mu e)/\sigma (\bar\nu_\mu e).$
The neutrinos were produced from muon decay and scattered from a fixed
target.  The ratio $R$ allows a determination of $\sin^2\theta_W ,$
while the two separate cross sections allow a determination of the $\rho$
parameter as well as the weak mixing angle.  In the $\overline{MS}$ scheme,
$\sin^2\hat\theta_W(M^2_Z)$ = 0.237$\pm$0.010(exp)$\pm$0.002(th), and
$\hat\rho$ = 1.001$\pm$0.038(exp)$\pm$0.004(th)~[4].  

In the on-shell definition, the combined weighted average of all experiments
(CHARM I, CHARM II, E734) yields: $\sin^2\theta_W$ = 0.224$\pm$0.009.
\bigskip
{\twelveit 5. $W$ and $Z$ Gauge Boson Properties}
\bigskip
The values of the $W$ and $Z$ gauge boson masses and widths are quoted here
in the on-shell renormalization scheme.

The measurement of the $W$ gauge boson properties from direct production has
been performed by the UA1~[15] and UA2~[16] collaborations of CERN and the 
CDF~[17] and D0~[18]
collaborations of Fermilab.  The combined world average of the $W$ mass
is $M_W$ = 80.23$\pm$0.18 GeV, with a combined world average $W$ width of
$\Gamma_W$ = 2.076$\pm$0.077 GeV~[19].  The $W$ width is measured by both
direct counting of decays and by examining ratios of partial widths of
$W$ to $Z$ decays, combined with $Z$ partial widths measured separately
(see below).

The next group of measurements of the $W$ boson properties are planned for
CDF and D0 (Fermilab) and LEPII (CERN).  The CDF/D0 program is one of
continuing improvement, starting from Run I (1992-93) of 75 $pb^{-1}$
to 200 $pb^{-1}$ by 1997.  With the Main Injector operating (scheduled
1998), the subsequent Run II is projected to accumulate several hundred
$pb^{-1}.$  More speculative is Run III, starting approximately in 2004,
accumulating perhaps as much as 5 $fb^{-1}.$  The projected error in $M_W$
by the end of Run II is 50 MeV, limited at that point by systematics, with
a comparable error in $\Gamma_W$~[20].  LEPII is scheduled to begin operations
by 1997 and to reach, by 2000, an accuracy in $M_W$ and $\Gamma_W$ of 
30-50 MeV~[21].

The bulk of the precision measurements of the $Z$ boson properties come
from the four experiments (ALEPH, DELPHI, OPAL, L3) at the LEP $e^+e^-$ 
collider at CERN, which has been in operation since 1989 and has accumulated
about eight million $Z$ events~[4,22,23].  The basic process is 
$e^+e^-\rightarrow Z,$ decaying to final-state pairs of leptons and quarks.  
The $Z$ mass and width are now measured to 91.189$\pm$0.004
GeV and 2.497$\pm$0.004 GeV, respectively.  The LEP collaborations have
measured a variety of asymmetries: forward-backward asymmetry $A_{FB}(Z)$
to leptonic and bottom and charm quark final states, and the $\tau$ and $e$
polarization asymmetries $A_{\tau,e} (Z)$ (asymmetry to left- and right-handed 
$\tau, e$).  In addition, the hadronic-to-leptonic
and the bottom- and charm-to-hadronic width ratios have been
measured, with implications for the strong coupling $\alpha_s(M^2_Z)$ and
virtual top quark effects in the $Z\rightarrow b\bar b$ vertex.  The
asymmetries are, in effect, measurements of an effective $\sin^2\theta_W,$
essentially the Kennedy-Lynn $s^2_*(Z)$~[5] plus weak vertex corrections:
$\sin^2\theta^{eff}_W$ = 0.2323$\pm$0.0002(exp)$\pm$0.0002(th).  A LEP-only
fit for $m_t$ yields $m_t$ = 181$\pm$14(exp)$\pm$20(th) GeV, including the
two-loop ${\cal O}(\alpha\alpha_s)$ QCD correction~[22], in good agreement with
the CDF direct production result, $m_t$ = 174$\pm$16 GeV~[24].  The
theoretical uncertainties are due to the unknown Higgs boson mass.  The
width measurements are all in agreement with the SM predictions, with the
exception of the $Z\rightarrow b\bar b$ partial width, which is 2.2$\sigma$
above the SM prediction.

The SLC $e^+e^-$ collider at SLAC has a unique ability to polarize its
electron beam and thus to measure the left-right polarization asymmetry
$A_{LR}(Z),$ the difference of $Z$ production with left- and right-handed
initial electrons.  The SLC collider commenced operations in 1989, but did
not begin polarized measurements until 1992.  Slightly fewer than 50,000
polarized events have been accumulated, with average polarization
(66$\pm$1)\%, but certain features of $A_{LR}(Z)$ compensate for this smaller 
data sample, in comparison with the LEP asymmetry measurements.  Through
1993, the SLD collaboration has measured $\sin^2\theta^{eff}_W$ =
0.2294$\pm$0.0010, which is 2.8$\sigma$ below the LEP result,
an as-yet unexplained discrepancy~[4,23,24].

LEP is scheduled to run until the end of 1995, with projected improvements
in: $\sin^2\theta_W$ to 0.0003, $\Gamma_Z$ to 2 MeV, and $\Gamma^{bb}_Z$ to
0.5\% .  A polarization measurement at LEP has been proposed for 1996, which,
if carried out, could yield a measurement of $\sin^2\theta_W$ to 0.0001~[26].
The next polarization run at SLC, starting in 1994, is projected to produce
100-150,000 polarized $Z$ events, with a consequent accuracy in $\sin^2\theta_W$
of 0.0005~[27].
\bigskip
{\twelveit 6. Hadronic Decay $Z\rightarrow b\bar b$}
\bigskip
This flavor-specific decay mode is, at the $Z$ pole, uniquely sensitive to
physics involving heavy fermion masses, as the $b$ quark is in the third
family.  The deviation of this quantity from its Standard Model value is not
in the same class as the $S,$ $T,$ $U$ corrections, because it is not universal
to all fermion final states of the $Z.$  Rather, the decay width depends
on the special $Z\rightarrow b\bar b$ vertex, which receives additional heavy
top quark mass corrections beyond the universal $T$ correction~[4,5].  The 
measurement from CERN/LEP is extracted from the branching ratio $R_b$ =
$\Gamma (Z\rightarrow b\bar b)/\Gamma (Z\rightarrow hadrons).$  This ratio is 
at present 1.8$\sigma$ above its Standard Model prediction.  With
$$\eqalign{
\Gamma (Z\rightarrow b\bar b) = \Gamma^0(Z\rightarrow b\bar b)(1 +
\gamma_b)\quad ,}
$$\noindent where $\Gamma^0(Z\rightarrow b\bar b)$ includes the universal $S,$
$T,$ $U,$ and $\rho_0$ corrections, we have, for non-Standard contributions:
$$\eqalign{
\gamma^{NS}_b = 0.032\pm 0.016\quad .}
$$\noindent In the SM,
$$\eqalign{
\gamma^{t\bar t}_b\simeq -(0.01)\big[ {m^2_t\over 2M^2_Z} - {1\over 5}\big]}
$$\noindent has already been removed to obtain the result $\gamma^{NS}_b$ by
using the CDF value of $m_t.$

This measurement can be used to place a $T-$ and $\rho-$independent constraint
on the top quark mass of $m_t$ = 175$\pm$16 GeV~[4].
\bigskip
{\twelveit 7. Rare $Z$ Gauge Boson Decays}
\bigskip
Decays of the $Z$ gauge boson with small branching ratios can be divided
into three categories.  The first consists of rare decays expected within
the Standard Model; the second, of strictly forbidden decays with only
Standard Model states; and the third, of decays with non-Standard particles.
The last class is necessarily as aspect of new, non-Standard particle
searches and is not discussed here.  Of the accelerators that have
produced $Z$ bosons (Sp$\bar{\rm p}$S, Tevatron, SLC, LEP), only LEP has
produced $Z$ events in numbers sufficient to make study of rare $Z$
decays possible.

The LEP collaborations have studied the Standard Model rare processes
$Z\rightarrow\gamma\gamma\gamma ,$ $Z\rightarrow\gamma + PS(V)$ (with
$PS(V)$ = pseudoscalar or vector meson),
$Z\rightarrow\gamma^*\nu\bar\nu$ (with $\gamma^*\rightarrow f\bar f$),
and $Z\rightarrow\gamma\gamma f\bar f$~[4,23].  
The first two searches have found 
rates consistent with the Standard Model.  The third search recorded events in 
excess of expectation, but still allowed at the 5\% level.  The status of final
search is unclear, as the L3 collaboration
found four such events with invariant $\gamma\gamma$ mass of about 60 GeV,
more than a factor of ten greater than the Standard Model expectation,
suggesting associated production of a state $X$ in the $Z$ decay, with
$X\rightarrow\gamma\gamma$ and $m_X\simeq$ 60 GeV.  However, subsequent
DELPHI searches have found no invariant $\gamma\gamma$ peak at this
energy and no inconsistency with the Standard Model.

The LEP experiments have also been used to search for the strictly
forbidden decays: $Z\rightarrow e\mu ,e\tau ,$ and $\mu\tau ,$ and
$Z\rightarrow pe$ (or $\mu$).  The first set violates the separate
lepton family numbers, which, in the absence of neutrino masses and
mixings, are exactly conserved.  The second violates baryon and lepton
numbers.  Both searches have found no evidence for such decays.
\bigskip
{\twelveit 8. Atomic Parity Violation}
\bigskip
A number of atomic parity violation
(APV) experiments before 1988 reported results for $\sin^2\theta_W;$ however, 
the completion of the \underbar{cesium} measurement at the University of 
Colorado, Boulder, of Wieman {\twelveit et al.} in
1988 raised the experimental precision of APV to a significantly higher level.
Their result for the atomic {\twelveit weak charge} $Q^{Cs}_W$ = $-$71.04$\pm$
1.58(exp)$\pm$0.88(th)~[28].  The atomic theory calculation was carried out by
Sapirstein {\twelveit et al.} in 1990~[29].  Combined with $M_Z$ measured at 
LEP, this APV measurement alone yields an unusually negative value of 
$S\simeq -3$~[30],
albeit with a large uncertainty $\simeq\pm$4.  The measurement is performed
with atomic transitions in crossed electric and magnetic fields.

Wieman's group at Boulder is proceeding with an improved \underbar{cesium} 
measurement,
whose goal is an uncertainty in $Q^{Cs}_W$ of $\pm$0.30$-$0.35, with a
plan comparable uncertainty in atomic theory, calculated again by
Sapirstein {\twelveit et al.}  The new experiment will
use laser-trapped cesium atoms and, by using different isotopes of cesium,
can eliminate some of the atomic theory uncertainty.  This effort is
scheduled to be completed in 3-4 years, yielding a reduced uncertainty in
$S$ measured by APV alone of $\simeq\pm$1~[31,32].

A more recent program of APV is being conducted by Norval Fortson {\twelveit et
al.} of the University of Washington, Seattle, based on optical rotation of
polarized light in lead and thallium vapors.  This group has published an APV
measurement in \underbar{lead} with $\pm$1\% accuracy, the best so far~[33].  
The atomic theory, unfortunately, has uncertainty of approximately 
$\pm$8\%~[34]; much 
of this uncertainty can be eliminated by the use of different lead isotopes, 
as planned by the Fortson collaboration in the coming years.  A better 
measurement has been done in \underbar{thallium}, for which a fairly precise 
calculation has been done, to $\pm$3\%.  The experimental uncertainty is again 
$\pm$1\%, yielding $S$ = -2.3$\pm$3.2; the final experimental uncertainty 
should reach approximately $\pm$0.5\%~[35].  Sapirstein {\twelveit et al.}~[32]
and Martensson-Pendrill {\twelveit et al.}~[36] are calculating the atomic 
theory for thallium.  The Fortson group is also beginning to explore 
the use of trapped single atomic ions for APV measurements; for example, 
Ba$^+,$ similar to cesium~[35].
\bigskip
{\twelveit Summary of Future Experiments}
\bigskip
Between this year (1994) and 2000, the subject of precision measurements
of electroweak gauge interactions will be refined into its probable final
form.  

\item{$\bullet$} The $W$ boson mass $M_W$ will be measured to approximately 
$\pm$50 MeV
by the CDF and D0 collaborations at the Tevatron (Fermilab), and to $\pm 30-
50$ MeV at LEPII (CERN).

\item{$\bullet$} The top quark mass will be measured to $\pm$5$-$10 GeV by CDF 
and D0~[37].

\noindent These two measurements require the Main Injector at Fermilab, which 
is scheduled to operate in 1998~[20]; while LEPII is scheduled to begin 
operations in 1996/97~[26].

\item{$\bullet$} The NuTeV collaboration's E815 experiment at Fermilab will 
have measured $\sin^2\theta_W$ to $\pm$0.0025 by 1998 by DIS~[11].  

\item{$\bullet$} The Boulder and Seattle APV
measurements will result in an $S$ measurement to approximately $\pm1-1.5.$

\item{$\bullet$} In the next two years, the SLD collaboration at SLC/SLAC will
complete their measurement of $A_{LR}(Z)$ with 3-500,000 polarized $Z$ events 
at approximately 80\% polarization, with a result for $\sin^2\theta_W$
accurate to $\pm$0.004~[27].  

\noindent If the top quark mass is known to better than $\pm$10 GeV, the 
electroweak SM can be tested with negligible uncertainty to the one-loop level 
of perturbation theory~[4,5].  Together with improved
measurements of flavor-mixing and CP violation, gauge interactions and the 
fermion mass matrix will be available in essentially
complete form.  Any deviations of measurements from the minimal SM, with
known top quark mass, will provide clear evidence for heavy particle states
beyond the $Z$ mass and place constraints on the possible realizations of
the Higgs sector.  The presently available data already disfavor additions
to the minimal SM, unless of a special type that produces small to no
effects below the $Z$ pole.  A supersymmetric Higgs sector, with the
superpartners of known particles, matches the data well~[4,5], while strongly
coupled Higgs sectors such as technicolor are difficult to accommodate.
However, real knowledge of the Higgs sector requires direct exploration with
very high energy accelerators.
\vfill\eject
{\twelvebf Electroweak Gauge Theory}
\bigskip
The electroweak Standard Model is a non-Abelian gauge theory based on
the gauge group SU(2)$_L\times$U(1)$_Y$~[5,9,10,38].  The gauge symmetry is 
broken by the Higgs sector,
leaving an unbroken Abelian gauge group U(1)$_Q,$ the basis of quantum
electrodynamics (QED) with its massless photon $(\gamma ).$  The other three
gauge bosons, the charged $W^{\pm}$ and neutral $Z^\circ ,$ are massive,
with masses of approximately 80 and 91 GeV/$c^2,$ respectively; these mediate
low-energy Fermi four-fermion weak interactions, such as beta decay and
neutrino-nucleon scattering.  The underlying gauge couplings, the SU(2)$_L$
$g$ and the U(1)$_Y$ $g^\prime ,$ are small, allowing perturbative expansion in
powers of the couplings as a solution of the quantum theory.  Non-perturbative
solutions of the electroweak gauge theory have also been investigated,
including the so-called {\twelveit sphalerons}~[39] (please see the report of
Working Group 6: Astroparticle Physics, Cosmology, and Unification), but
these are not relevant to accelerator experiments.

At tree (classical) level, the electroweak gauge theory has four parameters,
equivalent to $g$, $g^\prime ,$ $\vev ,$ and $\rho .$  The electromagnetic 
coupling $\alpha$ = $e^2/4\pi ,$ the Fermi weak decay
constant $G_F,$ the sine of the weak mixing angle $\sin^2\theta_W,$ and the 
weak gauge boson masses $M_W$ and $M_Z$ are then:
$$\eqalign{
\alpha^{-1}_{\rm em} &= 4\pi /e^2 = 4\pi [1/g^2 + 1/g^{\prime 2}]\quad ,\cr
\tan\theta_W &= g^\prime /g\quad ,\cr
G_F &= 1/\sqrt{2}\langle\phi\rangle^2\quad ,\cr
M^2_Z &= {\pi\alpha\over G_F\sin^2\theta_W\cos^2\theta_W}\quad ,\cr
\rho &= M^2_W/M^2_Z\cos^2\theta_W\quad .}
$$\noindent
If the Higgs VEV is due solely to SU(2)$_L$ doublets, then $\rho$ = 1 
automatically.
Once higher-order quantum or loop corrections are introduced, then the
theory must be renormalized, and a set of arbitrarily but consistently
defined parameters, a {\twelveit renormalization scheme,} must be introduced to 
replace the classical parameters.  The quantum or radiative corrections fall
into two categories, {\twelveit universal} and {\twelveit non-universal;} that 
is,
corrections that shift the value of the classical parameters without changing
the form of classical interactions versus corrections that do not respect
the classical form.  The first type of corrections have been investigated
in the work of Kennedy and Lynn~[5], and Degrassi and Sirlin~[40], and can be 
related
to standard renormalization schemes.  The second type of corrections varies
depending on the specific process in question.  All information concerning
electroweak interactions is currently derived from four-fermion processes.
Three renormalization schemes are commonly used, the {\twelveit on-shell,} the
{\twelveit modified minimal subtraction $(\overline{MS}),$} and the {\twelveit 
Lynn-Peskin-Stuart (LPS)} schemes.  The on-shell weak mixing angle is defined 
by
$$\eqalign{
\sin^2\theta_W = 1 - M^2_W/M^2_Z\quad ,}
$$\noindent
while the $\overline{MS}$ $\sin^2\hat\theta_W(M^2_Z)$ is a running,
scale-dependent weak mixing angle defined through the $\overline{MS}$
regularization method.  It is particularly convenient for comparing electroweak
measurements at differing energies and $Z$ pole measurements with models of
grand unification of electroweak and strong forces.
Please see the report of Working Group
1: Tests of the Electroweak Theory, for further details.  The calculation of
the perturbative quantum field theory of electroweak interactions has been
advanced and essentially completed over the last twenty years by many
workers, including: Veltman, 't Hooft, Taylor, Passarino, Marciano, Sirlin,
Lynn, Stuart, Hollik, Jegerlehner, Jadach, Berends, Kleiss, 
B.~F.~L.~Ward, Kennedy, Peskin, Takeuchi, and others.

For the purposes of investigating states and interactions beyond the SM,
the crucial property of electroweak gauge theory is that the full gauge
symmetry SU(2)$_L\times$U(1)$_Y$
is broken by the Higgs VEV~[5].  This is necessary for the
appearance of {\twelveit non-decoupled} radiative corrections, which do not
vanish as inverse powers of heavy particle masses $M^2,$ in a process of
a given momentum transfer $q^2,$ with $M^2$ $\gg$ $q^2.$  Such effects
scale as $M^n$ ($n$ = 0 or 2) or $\ln (M^2).$  A further necessary condition
is that the virtual effect break a tree-level global symmetry; these two
conditions together are sufficient to produce non-decoupled effects.  For
flavor-conserving or -diagonal processes, the relevant global symmetries
are the weak chiral group SU(2)$_L\times$SU(2)$_R$
and its vector subgroup SU(2)$_V.$  Three 
conventional parameters, usually represented
as $S,$ $T,$ and $U,$  summarize
completely universal non-decoupled effects in electroweak gauge interactions.
The first, $S,$ breaks the global SU(2)$_L\times$SU(2)$_R$
group, while the last two, $T$ and
$U,$ break the global SU(2)$_V$
subgroup.  In the case of a general, non-doublet Higgs sector, the
classical parameter $\rho$ replaces $T:$ $1 - \alpha T\rightarrow 1/\rho .$
Complete one-loop calculations of $S,$ $T,$ and $U$ have been performed for
the minimal SM, as well as for the minimal SUSY SM and many technicolor
theories.  Without special cancellations, non-Standard physics is expected
to contribute to these parameters at ${\cal O}(0.1 - 1),$ apart from group
theoretical factors.

A summary of classical parameters:
$$\eqalign{
\alpha^{-1}_{\rm em} &= 137.0359895\pm 0.0000061\quad ,\cr
G_\mu &= (1.16639\pm 0.00002)\times 10^{-5}\ {\rm GeV}\quad ,\cr
M_Z &= 91.189\pm 0.004\ {\rm GeV}\quad ,\cr
\rho_0 &\equiv 1\quad ,}
$$\noindent
in the minimal, doublet-only Higgs case; otherwise~[4,5]:
$$\eqalign{
\rho_0 &= M^2_W/\hat\rho M^2_Z\cos^2\hat\theta_W(M^2_Z)\quad ,}
$$\noindent
with $\hat\rho^{-1}\simeq 1 - \alpha T.$
A global fit to all current and relevant electroweak data yields, in the 
minimal SM with doublet-only Higgs VEVs (May 1994)~[4]:
$$\eqalign{
\sin^2\hat\theta_W(M^2_Z) &= 0.2317\pm 0.0004\quad ,\cr
\sin^2\theta_W &= 0.2242\pm 0.0012\quad ,\cr
m_t &= 173\pm 11\pm 18\ {\rm GeV}\quad ,}
$$\noindent
including the two-loop ${\cal O}(\alpha\alpha_s)$
QCD correction to $m_t,$ where the first uncertainty
is experimental and the second due to the unknown Higgs boson mass.  This
top quark mass value is in almost exact agreement with the CDF value, $m_t$ =
174$\pm$16 GeV.  The ${\cal O}(\alpha\alpha^2_s)$ threshold correction raises
$m_t$ by about $+3$ GeV.  For the SM with general, non-doublet Higgs VEVs:
$$\eqalign{
\sin^2\hat\theta_W(M^2_Z) &= 0.2318\pm 0.0005\quad ,\cr
m_t &= 170\pm 16\ {\rm GeV}\quad ,\cr
\rho_0 &= 1.0004\pm 0.0003\quad ,}
$$\noindent
where $m_t$ is determined mainly by $R_b.$
For the SM with $S,$ $T,$ $U$ and doublet-only Higgs VEVs:
$$\eqalign{
\sin^2\hat\theta_W(M^2_Z) &= 0.2314\pm 0.0004\quad ,\cr
m_t &= 175\pm 16\ {\rm GeV}\quad ,\cr
S &= -0.15\pm 0.28\quad ,\cr
T &= -0.08\pm 0.35\quad ,\cr
U &= -0.56\pm 0.61\quad ,}
$$\noindent
where the top quark mass here is determined by $R_b.$  The CDF measured value
of the top quark mass is the reference for $S$ = $T$ = $U$ = 0.  These fits 
lead to a picture consistent, within one standard deviation,
with the minimal SM (where $m_t$ = 174$\pm$16 GeV),
with little room for non-Standard physics.
Note in particular that the central value of $S$ is equal to zero within
one standard deviation, a significant change from the recent trend of $S$
measurements, which had been more negative~[41].
The best projected precision electroweak measurements will
require an uncertainty in $m_t$ of about 10 GeV to eliminate $m_t$ as a
significant source of uncertainty in radiative corrections analyses.

The effect of strong interactions in principle introduces uncertainties
into electroweak calculations.  Insofar as these effects are computable
using perturbative QCD, the strong gauge coupling $\alpha_s$ must be
known.  Two distinct measurements have been deduced from the LEP data, both
within the $\overline{MS}$ renormalization scheme and the mininal SM framework.
The first is derived from the hadronic branching ratio $R_{had}$ and the total
$Z$ width $\Gamma_Z,$ excluding $\Gamma_b;$ this yields
$\hat\alpha_s(M^2_Z)$ = 0.124$\pm$0.006.  The second is derived
from hadronic jet topologies from hadronic $Z$ decays, yielding: 
$\hat\alpha_s(M^2_Z)$ = 0.123$\pm$0.005.
There are still significant (1-3$\sigma$) discrepancies between these results 
and the same $\overline{MS}$ coupling inferred from certain lower-energy 
measurements (DIS and $b$ and $c$ meson properties); the
latter values are all lower than the LEP value.  In the minimal SM with
$S,$ $T,$ and $U,$ the strong coupling $\hat\alpha_s(M^2_Z)$ = 0.103$\pm$0.011
is considerably lower, because of the effect of $\gamma_b$ in $\Gamma(Z
\rightarrow b\bar b)$~[4].
In the case of the hadronic
contribution to the QED vacuum polarization (photon self-energy), the
effect of low-lying resonances can be incorporated using a dispersion
relation with the $e^+e^-\rightarrow hadrons$ data~[5,42].  In the case of
quark final states at LEP and SLC, QCD effects are included by perturbative
computation or by forming quantities, such as $A_{LR}(Z),$ that are
insensitive to strong interactions~[43].  In hadron colliders (Tevatron,
Sp$\bar{\rm p}$S, HERA), QCD effects are calculated using perturbation theory,
renormalization group techniques, and structure functions~[44].
\bigskip
{\twelvebf Status of Symmetries in Electroweak Gauge Interactions}
\bigskip
$C, P, CP:$ {\twelveit Discrete, Global:} Quantum electrodynamics is known
experimentally to conserve $C$ and $P$ separately.  The electroweak SM
connects $C$ and $P$ together in such a way that $CP$ is conserved
in electroweak gauge interactions, while
$C$ and $P$ are separately violated~[5,38].  $P$ violation is used in atomic
parity violation, polarized $e$-nucleus scattering, and the polarization
asymmetry $A_{LR}(Z).$  $C$ violation is measured by the forward-backward
asymmetries $A_{FB}(Z)$ at the $Z$ pole.  The $\sin^2\theta_W$ measurements
from the two asymmetries should, by $CP$ symmetry, be identical, and there
is no clear evidence at present that this is not so.  The 2.3$\sigma$
discrepancy between the LEP $A_{FB}(Z)$ and SLC $A_{LR}(Z)$ measurements
of $\sin^2\theta_W,$ if real, would be a signal of $CP$ violation in
electroweak gauge forces or of a new interaction.

$B ,L:$ {\twelveit Continuous, Global:} Baryon and lepton numbers are exactly
conserved in the SM, and there is no evidence of any $B$ or $L$ violation
at present~[4].  Quark mass mixing allows the transformation of baryons of
one family to another, while leaving $B$ fixed.  The separate $L$ family
numbers, however, are conserved.  The only evidence of lepton family mixing
at present comes from various possible signals of neutrino oscillations;
however, such effects are negligible in high-energy accelerator experiments.  
The electroweak gauge interactions
respect the separate family $B$ and $L$ quantum numbers, with no 
contrary experimental evidence.

SU(2)$_L\times$U(1)$_Y:$
{\twelveit Continuous, Local:} The electroweak gauge symmetry SU(2)$_L\times$U(1)$_Y$
accounts for the four electroweak gauge bosons and the relation of the
weak charged and neutral currents mediated by the massive $W$ and $Z$
gauge bosons.  Parity violation in both weak currents is predicted
correctly, as is the relation of the $W$ and $Z$ gauge boson masses
(see below).  This symmetry is broken by the Higgs
sector, with the U(1)$_Q$ gauge symmetry of QED left unbroken.  The exact
unbroken QED symmetry implies a massless photon and conserved electric
charge, both tested experimentally to high accuracy~[4].  There is no positive
evidence at present for a larger electroweak gauge group or new weak
gauge bosons~[4].

SU(2)$_L\times$SU(2)$_R:$
{\twelveit Continuous, Global:} The global weak chiral symmetry is
apparently an exact symmetry of the Higgs sector, broken spontaneously
by the Higgs VEV down to the weak chiral custodial subgroup SU(2)$_V$ (see
below).  Apart from the static Higgs VEV breaking of SU(2)$_L\times$SU(2)$_R,$
the radiative parameter $S$ measures the dynamical, momentum-dependent
breaking of SU(2)$_L\times$SU(2)$_R$ arising from loop corrections~[5].  
The present value of $S$ shows no significant deviation from zero.

SU(2)$_V:$
{\twelveit Continuous, Global:} The weak chiral custodial subgroup SU(2)$_V$
is the vector subgroup of SU(2)$_L\times$SU(2)$_R.$
It is respected by the Higgs sector to high accuracy, as measured by the 
parameter $\rho .$  The deviation from $\rho$ = 1 is accounted for the large 
top-bottom quark and the $Z-W$ mass splittings.  The residual deviation from
unity, indicating a general, non-doublet Higgs sector, is zero at 1.33
standard deviations.  The radiative parameters $T$ and $U$ measure the 
violation of SU(2)$_V$ by loop corrections, beyond the minimal SM
content~[5].  They are both zero within one standard deviation.

{\twelveit GIM Family Symmetry: Continuous, Global:} This symmetry rotates all
up-type quarks into one another and all down-type quarks into one another.
The electroweak gauge interactions respect this symmetry, preventing
flavor-changing neutral currents (FCNCs);the quark mass matrix does not.  This 
violation of GIM symmetry gives rise to FCNCs, but at the loop level only, and 
suppressed by $G^2_Fm^2_{q^\prime}m^2_q$~[5].  All
observed FCNCs $(K, D,$ and $B$ mesons) are consistent with
the minimal Cabibbo-Kobayashi-Maskawa quark mass matrix mixing~[4].

{\twelveit Hypothetical Symmetries:} These include {\twelveit new gauge groups}
and
{\twelveit gauge bosons,} {\twelveit supersymmetry,} and {\twelveit 
technicolor.}  The first
and last are local symmetries, while the second is global.  There is at
present no evidence for new weak gauge bosons, supersymmetric partners of
Standard Model states, or technifermions and technicolor gauge bosons~[4].
The last two would have radiative effects in weak interactions; supersymmetry
small to negligible, technicolor generally moderate to large.  The lack of
significant deviation in present data from the minimal SM tends to favor
supersymmetry, but only negatively, by the absence of any effect~[45].  
Consistent
technicolor theories respecting the electroweak and FCNC precision constraints
have yet to be constructed, although a several general schemes have been
proposed~[46].  Such technicolor theories would have to produce almost no
SU(2)$_V$ custodial breaking, beyond the top-bottom quark mass splitting, while
having either small technifermion sectors or special cancellations to guarantee
$S\simeq$ 1 or smaller.
\vskip 0.5in
{\bf Acknowledgements}
\bigskip
The author would like to thank for assistance and information: the Particle
Data Group, Lawrence Berkeley Laboratory, University of California, Berkeley;
27$^{th}$ International Conference on High Energy Physics, Glasgow (July 1994);
Robert Bernstein of CCFR (Fermilab);
Alain Blondel of CDHS and LEP/ALEPH (CERN);
Norval Fortson of the University of Washington, Seattle;
John Huth and Chris Wendt of CDF (Fermilab);
Paul Langacker of the University of Pennsylvania;
Bolek Pietrzyk of LEP/ALEPH (CERN);
Jonathan Sapirstein of Notre Dame University;
Morris Swartz of SLC/SLD (SLAC); and
Carl Wieman of the University of Colorado, Boulder.

This research was supported at the Institute for Theoretical
Physics, University of California, Santa Barbara, by the National Science 
Foundation under Grant No. PHY89-04035; by the University of Florida, Institute
for Fundamental Theory of the Department of Physics; and by the Department of 
Energy under Grant No. DE-FG05-86-ER40272.  The author would like to thank the 
ITP, where part of this study was prepared, for its hospitality.
\vskip 0.5in
{\bf References}
\bigskip
\item{ 1.} J.~M.~Jauch and F.~Rohrlich, {\twelveit The Theory of Photons and
Electrons} (Berlin: Springer-Verlag, 1976); V.~B.~Berestetskii, E.~M.~Lifshitz,
and L.~P.~Pitaevskii, {\twelveit Quantum Electrodynamics} (Oxford: Pergamon
Press, 1982).

\item{ 2.} T.~Kinoshita, ed., {\twelveit Quantum Electrodynamics} (Singapore: 
World Scientific, 1982).

\item{ 3.} T.~Kinoshita and A.~Sirlin, {\twelveit Phys. Rev.} {\twelvebf 113}
(1959) 1652.

\item{ 4.} Particle Data Group, {\twelveit Review of Particle Properties, Phys. 
Rev.} {\twelvebf D50}, Part~I (1994) 1173-1826: Gauge boson decays, 1191, 1351;
P.~Langacker and J.~Erler, "Standard Model of Electroweak Interactions," 1304,
and "Constraints on New Physics from Electroweak Analyses," 1312; P.~Langacker,
U. Pennsylvania preprint UPR-0624-T (1994).

\item{ 5.} D.~C.~Kennedy, {\twelveit Renormalization of Electroweak Gauge
Interactions,} in R.~K.~Ellis, C.~T.~Hill, and J.~D.~Lykken, eds., {\twelveit 
Perspectives in the Standard Model,} proc. 1991 TASI (Singapore: World 
Scientific, 1992) 163-280.

\item{ 6.} H.~Abramowicz {\twelveit et al.} (CDHS), {\twelveit Phys. Rev. Lett.}
{\twelvebf 57} (1986) 298; A.~Blondel {\twelveit at al.} (CDHS), {\twelveit
Z. Phys.} {\twelvebf C45} (1990) 361.

\item{ 7.} J.~V.~Allaby {\twelveit et al.} (CHARM), {\twelveit Z. Phys.}
{\twelvebf C36} (1987) 611.

\item{ 8.} C.~G.~Arroyo {\twelveit et al.} (CCFR), {\twelveit Phys. Rev. Lett.}
{\twelvebf 72} (1994) 3452.

\item{ 9.} C.~Quigg, {\twelveit Gauge Theories of the Strong, Weak, and
Electromagnetic Interactions} (Redwood City, CA: Addison-Wesley, 1983).

\item{10.} T.-P.~Cheng and L.-F.~Li, {\twelveit Gauge Theory of Elementary 
Particle Physics} (New York: Oxford University Press, 1984).

\item{11.} R.~Bernstein (CCFR/Fermilab), personal communication; {\twelveit
Fermilab Program Through 1997 and Beyond: Supplemental Material submitted to
the 1992 HEPAP Subpanel} (Batavia, IL: Fermilab, 1992) sec.~5.3.

\item{12.} J.~Dorenbosch {\twelveit et al.} (CHARM~I), {\twelveit Z. Phys.}
{\twelvebf C41} (1989) 567.

\item{13.} L.~A.~Ahrens {\twelveit et al.} (E734), {\twelveit Phys. Rev.}
{\twelvebf D41} (1990) 3297.

\item{14.} P.~Villain {\twelveit et al.} (CHARM~II), {\twelveit Phys. Lett.}
{\twelvebf B281} (1992) 159; {\twelvebf B335} (1994) 246.

\item{15.} G.~Arnison {\twelveit et al.} (UA1), {\twelveit Phys. Lett.} 
{\twelvebf 126B} (1983) 398.

\item{16.} J.~Alitti {\twelveit et al.} (UA2), {\twelveit Phys Lett.}
{\twelvebf 276B} (1992) 354.

\item{17.} H.~J.~Frisch {\twelveit et al.} (CDF), FERMILAB-CONF-94/44-E (1994);
Y.-K.~Kim {\twelveit et al.} (CDF), FERMILAB-PUB-94/169-E (1994). 

\item{18.} P.~Z.~Quintas {\twelveit et al.} (D0), FERMILAB-CONF-94/341-E (1994).

\item{19.} J.~Huth (CDF/Fermilab), personal communication.

\item{20.} C.~Wendt (CDF/Fermilab), personal communication.

\item{21.} J.~Ellis and R.~Peccei, eds., {\twelveit Physics at LEP,} CERN 86-02,(Geneva: CERN, 1986) vol.~2, 1.

\item{22.} B.~Pietrzyk, Laboratoire de Physique des Particules (LAPP) preprint
LAPP-EXP-94.07 (1994).

\item{23.} Proceedings, 27$^{th}$ International Conference on High Energy
Physics, Glasgow (1994), World Wide Web http://darssrv1.cern.ch/ichep.html1.

\item{24.} F.~Abe {\twelveit et al.} (CDF), {\twelveit Phys. Rev. Lett.}
{\twelvebf 73} (1994) 225.

\item{25.} K.~Abe {\twelveit et al.} (SLD), {\twelveit Phys. Rev. Lett.}
{\twelvebf 73} (1994) 25.

\item{26.} A.~Blondel (LEP/ALEPH), personal communication.

\item{27.} M.~Swartz (SLC/SLD), personal communication.

\item{28.} M.~C.~Noecker {\twelveit et al., Phys. Rev. Lett.} {\twelvebf 61}
(1988) 310.

\item{29.} S.~A.~Blundell {\twelveit et al., Phys. Rev.} {\twelvebf D45} (1992)
1602.

\item{30.} E.~R.~Boston and P.~G.~H.~Sandars, {\twelveit J. Phys.} {\twelvebf
B23} (1990) 2663; W.~Marciano and J.~L.~Rosner, {\twelveit Phys. Rev. Lett.}
{\twelvebf 65} (1990) 2963.

\item{31.} C.~E.~Wieman (U. Colorado/NIST), personal communication.

\item{32.} J.~Sapirstein (Notre Dame U.), personal communication.

\item{33.} D.~M.~Meekhof {\twelveit et al., Phys. Rev. Lett.} {\twelvebf 71} 
(1993) 3442; N.~Fortson (U. Washington, Seattle), personal communication.  

\item{34.} V.~A.~Dzuba {\twelveit et al., Z. Phys.} {\twelvebf D1} (1986) 243;
S.~J.~Pollock {\twelveit et al., Phys. Rev.} {\twelvebf C46} (1992) 2587;
N.~Fortson, J.~Sapirstein, personal communications.

\item{35.} V.~A.~Dzuba {\twelveit et al., Europhys. Lett.} {\twelvebf 7} (1988)
413; P.~Vetter {\twelveit et al., Bull. Am. Phys. Soc.} {\twelvebf 38} (1993)
1121; N.~Fortson, {\twelveit Phys. Rev. Lett.} {\twelvebf 70} (1993) 2383;
N.~Fortson, J.~Sapirstein, personal communications.

\item{36.} S.~J.~Pollock and N.~Fortson, U.~Washington preprint 
DOE-ER-40427-09-N92 (1992); N.~Fortson, personal communication.

\item{37.} Ref.~[11], {\twelveit Fermilab Program,} sec.~3.2.

\item{38.} J.~F.~Gunion {\twelveit et al., The Higgs Hunter's Guide} (Redwood
City, CA: Addison-Wesley, 1990); M.~B.~Einhorn, ed., {\twelveit The Standard 
Model Higgs Boson} (Amsterdam: North-Holland, 1991); P.~Langacker, ed., 
{\twelveit Precision Tests of the Standard Electroweak Model} (Singapore: World
Scientific, 1993).

\item{39.} P.~B.~Arnold, {\twelveit An Introduction to Baryon Violation in
Standard Electroweak Theory,} in M.~Cveti\v c and P.~Langacker, eds., 
{\twelveit Testing the Standard Model,} proc. 1990 TASI (Singapore: World
Scientific, 1991) 719-742.

\item{40.} G.~Degrassi and A.~Sirlin, {\twelveit Nucl. Phys.} {\twelvebf B383}
(1992) 73; {\twelveit Phys. Rev.} {\twelvebf D46} (1992) 3104.

\item{41.} D.~C.~Kennedy, FERMILAB-CONF-93/23-T (1993), to be published in
proc. 21$^{st}$ Coral Gables/Global Foundation Conference (New York: Nova
Science, 1994); P.~Langacker, {\twelveit Precision Tests of the Standard Model,}
in J.~Harvey and J.~Polchinski, eds., {\twelveit From Superstrings and Black 
Holes to the Standard Model,} proc. 1992 TASI (Singapore: World Scientific, 
1993) 141-162.

\item{42.} H.~Burkhardt {\twelveit et al.,} in G.~Alexander, ed., {\twelveit
Polarization at LEP,} CERN 88-06 (Geneva: CERN, 1986) vol~2, 145.

\item{43.} T.~Muta, {\twelveit Foundations of Quantum Chromodynamics}
(Singapore: World Scientific, 1987).

\item{44.} R.~Field, {\twelveit Applications of Perturbative QCD} (Redwood
City, CA: Addison-Wesley, 1989).

\item{45.} H.~P.~Nilles, {\twelveit Phys. Rep.} {\twelvebf 110C} (1984) 1;
H.~E.~Haber and G.~Kane, {\twelveit Phys. Rep.} {\twelvebf 117C} (1985) 76;
B.~W.~Lynn, SLAC preprint SLAC-PUB-3358 (1984); R.~Barbieri 
{\twelveit et al., Nucl. Phys.} {\twelvebf B341} (1990) 309. 

\item{46.} S.~Weinberg, {\twelveit Phys. Rev.} {\twelvebf D19} (1979) 1277;
L.~Susskind, {\twelveit Phys. Rev.} {\twelvebf D20} (1979) 2619;
S.~Dimopoulos and L.~Susskind, {\twelveit Nucl. Phys.} {\twelvebf B155} (1979)
237; E.~Eichten and K.~Lane, {\twelveit Phys. Lett.} {\twelvebf 90B} (1980) 
125; T.~Appelquist {\twelveit et al., Phys. Rev. Lett.} {\twelvebf 57} (1986) 
957; B.~Holdom, {\twelveit Phys. Lett.} {\twelvebf 198B} (1987) 535;
M.~Golden and L.~Randall, {\twelveit Nucl. Phys.} {\twelvebf B361} (1991) 3;
R.~Sundrum and S.~D.~H.~Hsu, {\twelveit Nucl. Phys.} {\twelvebf B391} (1993)
127; T.~Appelquist and G.~Triantaphyllou, {\twelveit Phys. Lett.} {\twelvebf
278B} (1992) 345; W.~A.~Bardeen {\twelveit et al., Phys. Rev.} {\twelvebf D41}
(1990) 1647; W.~J.~Marciano, {\twelveit Phys. Rev.} {\twelvebf D41} (1990) 219;
E.~Gates and J.~Terning, {\twelveit Phys. Rev. Lett.} {\twelvebf Phys. Rev.
Lett.} {\twelvebf 67} (1991) 1840; S.~Bertolini and A.~Sirlin, {\twelveit
Phys. Lett.} {\twelvebf 257B} (1991) 179; M.~Dugan and L.~Randall, {\twelveit
Phys. Lett.} {\twelvebf 264B} (1991) 154.

\vfill\eject
\end